\documentclass[journal,twocolumn]{IEEEtran}
\usepackage{amsfonts}
\usepackage{times}
\usepackage{graphicx}
\usepackage{latexsym}
\usepackage{dsfont}
\usepackage{amssymb}
\usepackage{amsmath}
\usepackage{cite}
\usepackage{verbatim}
\usepackage{subfigure}

\newcommand{\figref}[1]{{Fig.}~\ref{#1}}

\def\bb0{{\mathbb{0}}}

\def\ba{{\mathbf{a}}}
\def\bb{{\mathbf{b}}}

\def\bh{{\mathbf{h}}}

\def\bm{{\mathbf{m}}}

\def\bv{{\mathbf{v}}}

\def\by{{\mathbf{y}}}

\def\b0{{\mathbf{0}}}

\def\bA{{\mathbf{A}}}
\def\bB{{\mathbf{B}}}

\def\bF{{\mathbf{F}}}

\def\bH{{\mathbf{H}}}
\def\bI{{\mathbf{I}}}

\def\bP{{\mathbf{P}}}
\def\bQ{{\mathbf{Q}}}
\def\bR{{\mathbf{R}}}

\def\bU{{\mathbf{U}}}
\def\bV{{\mathbf{V}}}
\def\bW{{\mathbf{W}}}

\def\bY{{\mathbf{Y}}}

\def\bbE{{\mathbb{E}}}

\def\cA{\mathcal{A}}

\def\cN{\mathcal{N}}

\def\sf0{{\mathsf{0}}}

\def\vec{\mathrm{vec}~}
\def\kron{\otimes}

\usepackage{epstopdf}
\usepackage{enumerate}
\usepackage{algorithmicx}
\usepackage{algorithm}
\usepackage{amsmath}
\usepackage[noend]{algpseudocode}
\usepackage{float}
\usepackage{hyperref}
\usepackage{color}
\usepackage{makeidx}
\usepackage{bbm}
\usepackage{graphicx}
\usepackage{balance}

\newcommand{\sref}[1]{{Section}~\ref{#1}}

\DeclareMathOperator*{\argmax}{argmax} 

\begin{document}
\title{Deep Learning for Direct Hybrid Precoding in Millimeter Wave Massive MIMO Systems}
\author{Xiaofeng Li and Ahmed Alkhateeb\\ Arizona State University, Email:  $\left\{ \text{xiaofen2, alkhateeb} \right\}$@asu.edu}
\maketitle

\begin{abstract}
	This paper proposes a novel neural network architecture, that we call an \textit{auto-precoder}, and a deep-learning based approach that jointly senses the millimeter wave (mmWave) channel and designs the hybrid precoding matrices with only a few training pilots. More specifically, the proposed machine learning model leverages the prior observations of the channel to achieve two objectives. First, it optimizes the compressive channel sensing vectors based on the surrounding environment in an unsupervised manner to focus the sensing power on the most promising spatial directions. This is enabled by a novel neural network architecture that accounts for the constraints on the RF chains and models the transmitter/receiver measurement matrices as two complex-valued convolutional layers. Second, the proposed model learns how to construct the RF beamforming vectors of the hybrid architectures directly from the projected channel vector (the received sensing vector). The auto-precoder neural network that incorporates both the channel sensing and beam prediction is trained end-to-end as a multi-task classification problem. 	
	Thanks to this design methodology that leverages the prior channel observations and the implicit awareness about the surrounding environment/user distributions, the proposed approach  significantly reduces the training overhead compared to classical (non-machine learning) solutions. For example, for a system of 64 transmit and 64 receive antennas, with 3 RF chains at both sides, the proposed solution needs only 8 or 16 channel training pilots to directly predict the RF beamforming/combining vectors of the hybrid architectures and achieve near-optimal achievable rates. This highlights a promising solution for the channel estimation and hybrid precoding design problem in mmWave and massive MIMO systems.
	
\end{abstract}

\section{Introduction} \label{sec:Intro}
Hybrid analog/digital architectures have attracted significant interest in the last few years thanks to their capability of achieving high data rates with energy-efficient hardware. 
To design the hybrid precoding matrices, however, an explicit estimation of the mmWave channel is normally required. This mmWave channel estimation is a challenging task because of the large numbers of antennas at both the transmitters and receivers, which result in high training overhead, and the strict hardware constraints on the RF chains \cite{HeathJr2016,Alkhateeb2014d}. Leveraging the sparsity of the mmWave channels, several compressive sensing based channel estimation solutions have been proposed and showed promising performance \cite{Alkhateeb2014,HeathJr2016,Schniter2014,Lee2014}.  Essentially, these compressive sensing solutions for the mmWave channel estimation problem normally require an order of magnitude less training pilots compared to exhaustive search approaches \cite{Alkhateeb2015}. But can we do better? In this paper, we show that machine learning tools can efficiently leverage the prior observations about the channel estimates and the hybrid precoding designs to significantly reduce the training overhead associated with the mmWave channel training and precoding design problem.

Several channel estimation and hybrid precoding design approaches have been proposed in the last few years \cite{Alkhateeb2014,HeathJr2016,Schniter2014,Lee2015}. Most of these approaches relied on leveraging the sparse nature of the mmWave channels and developed compressive sensing based solutions for the channel estimation. The developed solutions in \cite{Alkhateeb2014,HeathJr2016,Schniter2014,Lee2015}  generally show that compressive measurements/projections can efficiently sense and reconstruct the mmWave channels while requiring less training pilots compared to the exhaustive beam training techniques.  These compressive sensing solutions, however, typically adopt random channel measurement vectors that distribute the sensing power in all spatial directions. For a given environment (for example, an outdoor street or an indoor room setting), though, one would expect that the base station/access point should focus the sensing power on the spatial directions that are likely to include the  angles or arrival/departure of the channel paths. Intuitively, these directions depend on the given environment (geometry, materials, etc.) and the candidate user locations among other factors.  Therefore, it is interesting to leverage any side information about the surrounding environment and user distribution in designing these sensing vectors. 

In this paper, we propose a novel deep-learning based approach the jointly optimizes the channel measurement vectors  and designs the hybrid beamforming vectors to achieve near-optimal data rates with negligible training overhead. More specifically, we develop a novel neural network architecture, that we call an \textit{auto-precoder}, to achieve two main objectives: (i) It learns how to optimize the channel sensing vectors to focus the sensing power on the promising spatial directions (which implicitly adapt these measurement vectors to the surrounding environment/ user distributions) and (ii) it learns  how to predict the hybrid beamforming vectors directly from the receive sensing vector (without the need to explicitly estimate/reconstruct the channel).  Achieving these two objectives results in a promising channel sensing/precoding design  approach that can predict near-optimal hybrid beamforming vectors while requiring negligible training overhead.

\textbf{Notation}: We use the following notation throughout this paper: $\bA$ is a matrix, $\ba$ is a vector, $a$ is a scalar, and $\cA$ is a set. $|\bA|$ is the determinant of $\bA$, whereas $\bA^T$ and $\bA^\mathsf{H}$ are its transpose and Hermitian (conjugate transpose).  $\bA \otimes \bB$ is the Kronecker product of $\bA$ and $\bB$, and $\bA \circ \bB$ is their Khatri-Rao product. $\cN_\mathbb{C}(\bm,\bR)$ is a complex Gaussian random vector with mean $\bm$ and covariance $\bR$. $\bbE\left[\cdot\right]$ is used to denote expectation.

\clearpage
\begin{figure}[t]
	\centerline{
		\includegraphics[width=1\columnwidth]{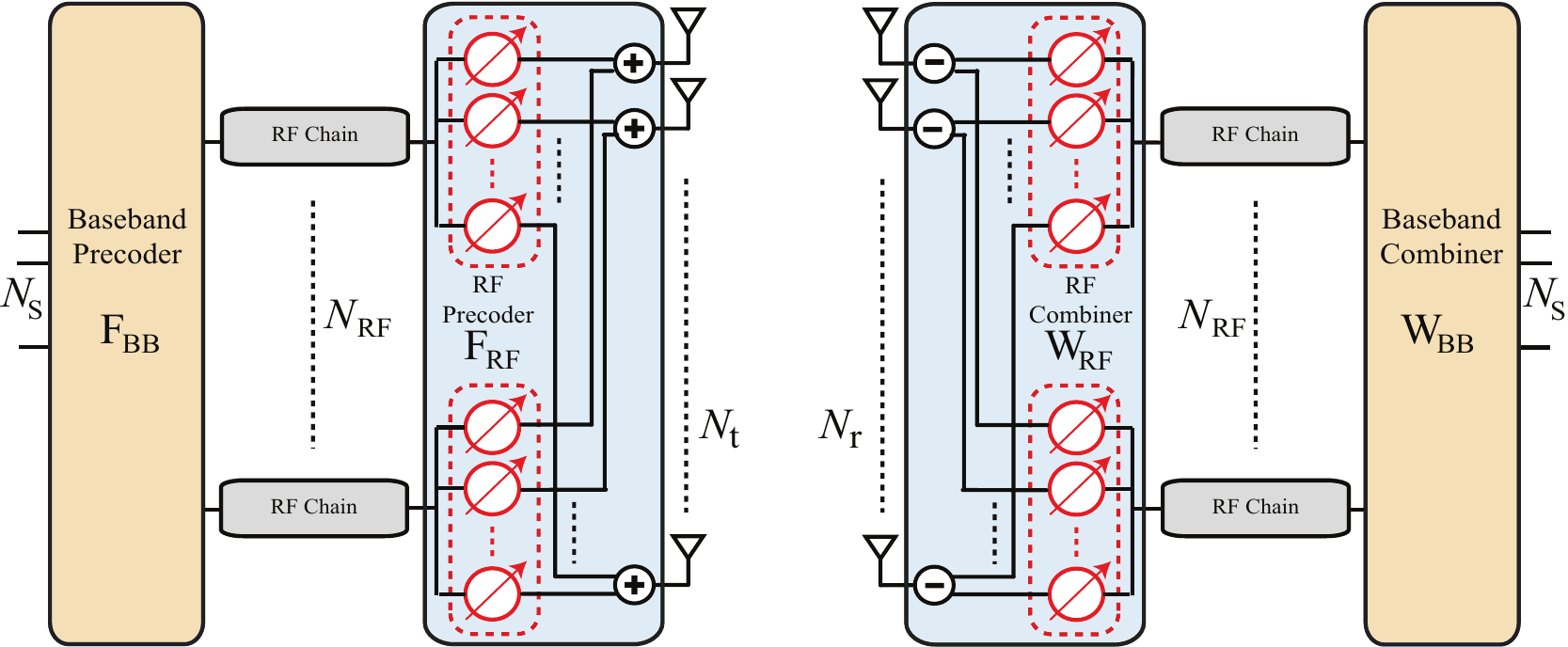}
	}
	\caption{A block diagram for the adopted hybrid analog/digital transceiver architecture with a transmitter employing $N_\mathrm{t}$ antennas  and $N_\mathrm{t}^\mathrm{RF}$ RF chains and a receiver employing $N_\mathrm{r}$ antennas  and $N_\mathrm{r}^\mathrm{RF}$ RF chains. The precoding/combining processing is divided between baseband precoding/combining $\bF_\mathrm{BB}, \bW_\mathrm{BB}$ and RF precoding/combining $\bF_\mathrm{RF}, \bW_\mathrm{RF}$. }
	\label{fig:hybrid}
\end{figure}

\section{System and Channel Models} \label{sec:Sys_Model}

We consider the fully-connected hybrid analog/digital architecture depicted in \figref{fig:hybrid}, where a transmitter  employing $N_\mathrm{t}$ antennas and $N^\mathrm{RF}_\mathrm{t}$ RF chains is communicating via $N_\mathrm{S}$  streams with a receiver having $N_\mathrm{r}$ antennas and $N^\mathrm{RF}_\mathrm{r}$ RF chains. The transmitter precodes the transmitted signal using an $N^\mathrm{RF}_\mathrm{t} \times N_\mathrm{S}$ baseband precoder $\bF_\mathrm{BB}$ and an $N_\mathrm{t} \times N^\mathrm{RF}_\mathrm{t}$ RF precoder $\bF_\mathrm{RF}$ while the receiver combines the received signal using the $N_\mathrm{r} \times N^\mathrm{RF}_\mathrm{r}$ RF combiner $\bW_\mathrm{RF}$ and the $N_\mathrm{r}^\mathrm{RF} \times N_\mathrm{S}$ baseband combiner $\bW_\mathrm{BB}$.  Since the analog/RF precoders/combiners, $\bF_\mathrm{RF}, \bW_\mathrm{RF}$ are implemented in the analog domain using RF circuits, every entry of the RF precoders/combiners is assumed to have a constant-modulus, i.e., $\left[\bF_\mathrm{RF}\right]_{n,m}=\frac{1}{\sqrt{N_\mathrm{t}}} e^{j \Theta_{m,nt}}$ (and similarly for elements of $\bW_\mathrm{RF}$). Further, the total power constraint is enforced by normalizing the baseband precoder $\bF_\mathrm{BB}$ to satisfy $\left\|\bF_\mathrm{RF} \bF_\mathrm{BB}\right\|_F^2=N_\mathrm{S}$ \cite{Alkhateeb2014}.

For the channel between the transmit and receive antennas, we adopt the geometric channel model in  \cite{Alkhateeb2014} with $L$ paths. In this model, the $N_\mathrm{r} \times N_\mathrm{t}$ channel matrix $\bH$ is written as 
\begin{equation}
\bH= \sum_{\ell=1}^L \alpha_\ell \ba_\mathrm{r}\left(\phi_{\mathrm{r},\ell},\theta_{\mathrm{r},\ell}\right) \ba^\mathsf{H}_\mathrm{t}\left(\phi_{\mathrm{t},\ell},\theta_{\mathrm{t},\ell}\right) 
\end{equation} 
where $\alpha_\ell$ denotes the complex path gain of the $\ell$th path (including the path loss). The angles $\phi_{\mathrm{r},\ell},\theta_{\mathrm{r},\ell}$ represent the $\ell$th path azimuth and elevation angles of arrival (AoAs) at the receive antennas while $\phi_{\mathrm{t},\ell},\theta_{\mathrm{t},\ell}$ represent the $\ell$th path angles of departure from the transmit array. Finally, $\ba_\mathrm{t}(.)$ and $\ba_\mathrm{r}(.)$ denote the transmit and receive array response vectors.  The array response vectors for ULA and UPA arrays are  defined in \cite{ElAyach2014}. It is worth noting here that for mmWave frequencies, the measurements showed that the channels are typically sparse in the angular domain resulting in a  small number of channel paths $L$ (normally in the range of 3-5 paths)\cite{Rappaport2013a,MacCartney2014}. 

\section{Problem Definition}
The general objective of this paper is to directly design the hybrid precoders/combiners to maximize the system achievable rate while minimizing the channel training overhead. Given the system and channel models in \sref{sec:Sys_Model}, the achievable rate with the analog/digital precoders/combiners can be written as 
\begin{equation}
R=\log_2\left|\bI+\bR_n^{-1} \bW^\mathsf{H} \bH \bF \bF^\mathsf{H} \bH^\mathsf{H} \bW\right|,
\label{eq:rate}
\end{equation}
with $\bF=\bF_\mathrm{RF} \bF_\mathrm{BB}$ and $\bW=\bW_\mathrm{RF} \bW_\mathrm{BB}$. The  matrix $\bR_n=\frac{1}{\mathsf{SNR}} \bW^\mathsf{H} \bW$ represents the noise covariance matrix where $\mathsf{SNR}=\frac{P_T}{N_\mathrm{S} \sigma_n^2}$, and with $P_T$ and $\sigma_n^2$ denoting the total transmit power and noise power.

Next, we assume that the RF beamforming/combining vectors are selected from pre-defined quantized codebooks, i.e., $\left[\bF_\mathrm{RF} \right]_{:,n_\mathrm{t}} \in\boldsymbol{\mathcal{F}}, \forall n_\mathrm{t}$ and $\left[\bW_\mathrm{RF} \right]_{:,n_\mathrm{r}} \in \boldsymbol{\mathcal{W}}, \forall n_\mathrm{r}$. If the channel is known, the hybrid precoder/combiner design problem can then be written as 
\begin{align}
\left\{\bF^\star_\mathrm{BB}, \bF^\star_\mathrm{RF} ,\bW^\star_\mathrm{BB}, \bW^\star_\mathrm{RF}\right\} = \hspace{10pt}& \nonumber\\
& \hspace{-90pt}\arg\max \hspace{12pt} \log_2\left|\bI+\bR_n^{-1} \bW^\mathsf{H} \bH \bF \bF^\mathsf{H} \bH^\mathsf{H} \bW\right|, \\
& \hspace{-87pt} \text{s.t} \hspace{38pt}   \bF=\bF_\mathrm{BB}\bF_\mathrm{RF}, \\
& \hspace{-42pt} \bW=\bW_\mathrm{BB} \bW_\mathrm{RF}, \\
& \hspace{-42pt} \left[\bF_\mathrm{RF} \right]_{:,n_\mathrm{t}} \in\boldsymbol{\mathcal{F}}, \forall n_\mathrm{t}, \\
& \hspace{-42pt} \left[\bW_\mathrm{RF} \right]_{:,n_\mathrm{r}} \in \boldsymbol{\mathcal{W}}, \forall n_\mathrm{r}, \\
& \hspace{-42pt} \left\|\bF_\mathrm{RF} \bF_\mathrm{BB}\right\|_F^2=N_\mathrm{S},
\end{align}

Further, if the RF beamforming./combining codebooks consist of orthogonal vectors (such as the case in the widely-adopted DFT codebooks), then for any selected RF precoders and combiners, $\bF_\mathrm{RF}, \bW_\mathrm{RF}$, the optimal baseband precoders/combiners are defined as \cite{Alkhateeb2016d}
\begin{align}
\bF_\mathrm{BB}^\star & = \left(\bF_\mathrm{RF}^\mathsf{H} \bF_\mathrm{RF}\right)^{-\frac{1}{2}} \left[\overline{\bV}\right]_{:,1:N_\mathrm{S}}, \label{eq:baseband1} \\
\bW_\mathrm{BB}^\star & = \left[\overline{\bU} \right]_{:,1:N_\mathrm{S}},
\label{eq:baseband2}
\end{align} 
where $\overline{\bV}$ and $\overline{\bU}$ are the right and left singular vector matrices of the effective channel matrix $\overline{\bH}=\bW_\mathrm{RF}^\mathsf{H} \bH \bF_\mathrm{RF}$. This reduces the hybrid precoding design problem to the following exhaustive search problem over the RF precoding/combining matrices
\begin{align}
& \left\{\bF^\star_\mathrm{RF}, \bW^\star_\mathrm{RF}\right\}  = \hspace{-7pt} \argmax_{\substack{\left[\bF_\mathrm{RF} \right]_{:,n_\mathrm{t}} \in\boldsymbol{\mathcal{F}}, \forall n_\mathrm{t} \\ \left[\bW_\mathrm{RF} \right]_{:,n_\mathrm{r}} \in \boldsymbol{\mathcal{W}}, \forall n_\mathrm{r}}} 
\hspace{-7pt} \log_2\left|\bI+\mathsf{SNR} \bW_\mathrm{RF}^\mathsf{H} \bH \bF_\mathrm{RF} \vphantom{\left(\bF_\mathrm{RF}^\mathsf{H} \bF_\mathrm{RF}\right)^{-1}}\right. \nonumber\\ 
& \hspace{120pt} \left. \times \left(\bF_\mathrm{RF}^\mathsf{H} \bF_\mathrm{RF}\right)^{-1} \bF_\mathrm{RF}^\mathsf{H} \bH^\mathsf{H} \bW_\mathrm{RF}\right|,
\label{eq:opt_hybrid}
\end{align}

The optimization problem in \eqref{eq:opt_hybrid} implies that the optimal hybrid precoders can be found via an exhaustive search over the candidate RF beamforming/combining vectors. The challenge however is that the channel is normally unknown and its explicit estimation requires very large training overhead in mmWave systems \cite{HeathJr2016}. To address this challenge, our objective in this paper is to devise a solution that directly finds the RF precoding/combining vectors of the hybrid architecture that maximize (or approach) the optimal achievable rate while requiring low channel training overhead. In the following section, we show how machine/deep learning tools can provide an efficient solution to this problem. 

\begin{figure*}[t]
	\centerline{
		\includegraphics[scale=2]{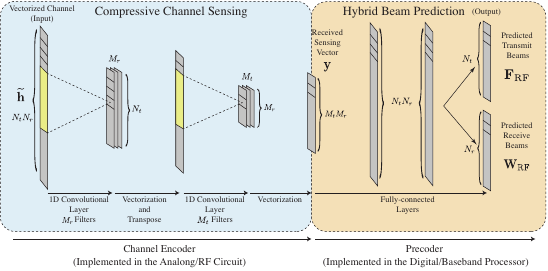}
	}
	\caption{The proposed \textit{auto-precoder} neural network consists of two sections: the channel encoder and the precoder. The channel encoder takes the vectorized channel vector as an input and passes it through two complex-valued 1D-convolution layers that mimic the Kronecker product operation of the transmit and receive measurement matrices in \eqref{eq:sensing}. The output of the channel encoder (the received sensing vector) is then inputted to the precoder network. The precoder network consists of  two fully connected layers and two output layers with the objective of predicting the indices of the $N_\mathrm{t}^\mathrm{RF}$ RF beamforming vectors and $N_\mathrm{r}^\mathrm{RF}$ RF combining vectors of the hybrid architecture. The dimensions of the two output layers are equal to the sizes of the RF beamforming and combining codebooks, $\boldsymbol{\mathcal{\bF}}, \boldsymbol{\mathcal{\bW}}$, i.e. the number of classes for the multi-label classification problem. }
	\label{fig:multi_label}
\end{figure*}

\section{mmWave Auto-Precoder: A Novel Neural Network for Direct Hybrid Precoding Design} \label{sec:hybrid}
In classical (non machine learning) signal processing, the channel estimation and hybrid precoding design is normally done through three stages \cite{Alkhateeb2014,Alkhateeb2014d}. First, leveraging the sparse nature of the mmWave channels, the channel is sensed using compressive measurements (that are normally random) \cite{Alkhateeb2014,HeathJr2016}. Then, the  channel is reconstructed from the compressed measurements using, for example, basis pursuit algorithms. Finally, the constructed channel is used to design the hybrid precoding matrices. The main drawback of this approach is that it does not leverage the prior channel observations to reduce the training overhead associated with estimating the channels and designing the precoders. In this paper, we propose a novel neural network architecture that we call an `auto-precoder' and a deep-learning approach that senses/compresses the channels and directly design the hybrid beamforming vectors from the compressed measurements. Our approach achieves multi-fold gains: (i) Different from the random measurements that are normally adopted in compressive channel estimation, our approach learns how to optimize the measurement (compressive sensing) vectors based on the user distribution and the surrounding environment to focus the measurement/sensing power on the most promising spatial directions. (ii) The deep learning model learns (and memorizes) how to predict the hybrid beamforming vectors directly from the compressed measurements. This design approach significantly reduces the training overhead while achieving near-optimal achievable rates as will be shown in \sref{sec:Results}. Next, we briefly explain the main concept of the proposed \textit{auto-precoder} deep learning model in \sref{subsec:Auto} and then provide a detailed description of its two main stages, namely the channel sensing and hybrid beam prediction in Sections \ref{sec:sensing}-\ref{subsec:Precode}.

\subsection{mmWave Auto-Precoder: The Main Concept} \label{subsec:Auto}

In this paper, we propose the novel \textit{auto-precoder} neural network architecture in \figref{fig:multi_label}. The auto-precoder neural network consists of two main sections: (i) the \textit{channel encoder} which learns how to optimize the compressive sensing vectors to focus the sensing power on the most promising directions and (ii) the \textit{precoder} which learns how to  predict the RF beamforming/combining vectors of the hybrid architecture directly from the receive sensing vector; i.e., the output of the channel encoder. In order to achieve these objectives, the auto-precoder network is trained and used as follows.
\begin{itemize}
	\item \textbf{Auto-Precoder Training:} In the training phase, the auto-precoder is trained end-to-end in a supervised manner. More specifically, a dataset of the mmWave channels and the corresponding RF beamforming/combining matrices are constructed and the auto-precoder is trained to be able to predict the indices of the RF precoding/combining vectors of the hybrid architecture given the input channel vector. In this paper, we use the near-optimal Gram-Schmidt hybrid beamforming algorithm in \cite{Alkhateeb2016d} to construct the RF beamforming/combining matrices. We note, however, that other hybrid beamforming algorithms can be also adopted to construct the target precoders. It is important to mention here that through the end-to-end training of the auto-precoder model, the channel encoder in \figref{fig:multi_label} learns in an unsupervised way how to optimize its compressive sensing vectors. This is thanks to the novel design of the channel encoder architecture that will be described in detail in \sref{sec:sensing}. 
	
	\item \textbf{Auto-Precoder Prediction:} After the auto-precoder network is trained, it is decoupled into two parts in the testing (prediction) phase: (i) the neural network of the channel encoder is directly implemented in the analog/RF circuits. More specifically, the weights of the two convolutional layers of the channel encoder network will be used as the weights of the analog/RF measurement matrices at both the transmitter and receiver.  This is enabled by the specific design of the channel encoder network as will be explained in \sref{sec:sensing}. These deep-learning optimized measurement matrices will be employed to sensing the unknown mmWave channel matrix. (ii) the output of the channel measurement, i.e., the receive sensing vector, will be inputted to the precoder network in the baseband/digital domain and used to directly predict the indices of the RF beamforming/combining vectors of the hybrid architecture. 
\end{itemize}

It is worth mentioning here that we focus in this paper on predicting the RF precoding/combining matrices of the hybrid architecture as their design is what normally requires large training overhead. Once the RF precoders/combiners are designed, the effective channel will have small dimensions, $N_\mathrm{r}^\mathrm{RF} \times N_\mathrm{t}^\mathrm{RF}$, and can be easily trained with a few training pilots to design the baseband precoders/combiners. In the next two sections, we will describe the two components of the auto-precoders model in \figref{fig:multi_label}.

\begin{figure}
	\centering
	\includegraphics[width=1\columnwidth]{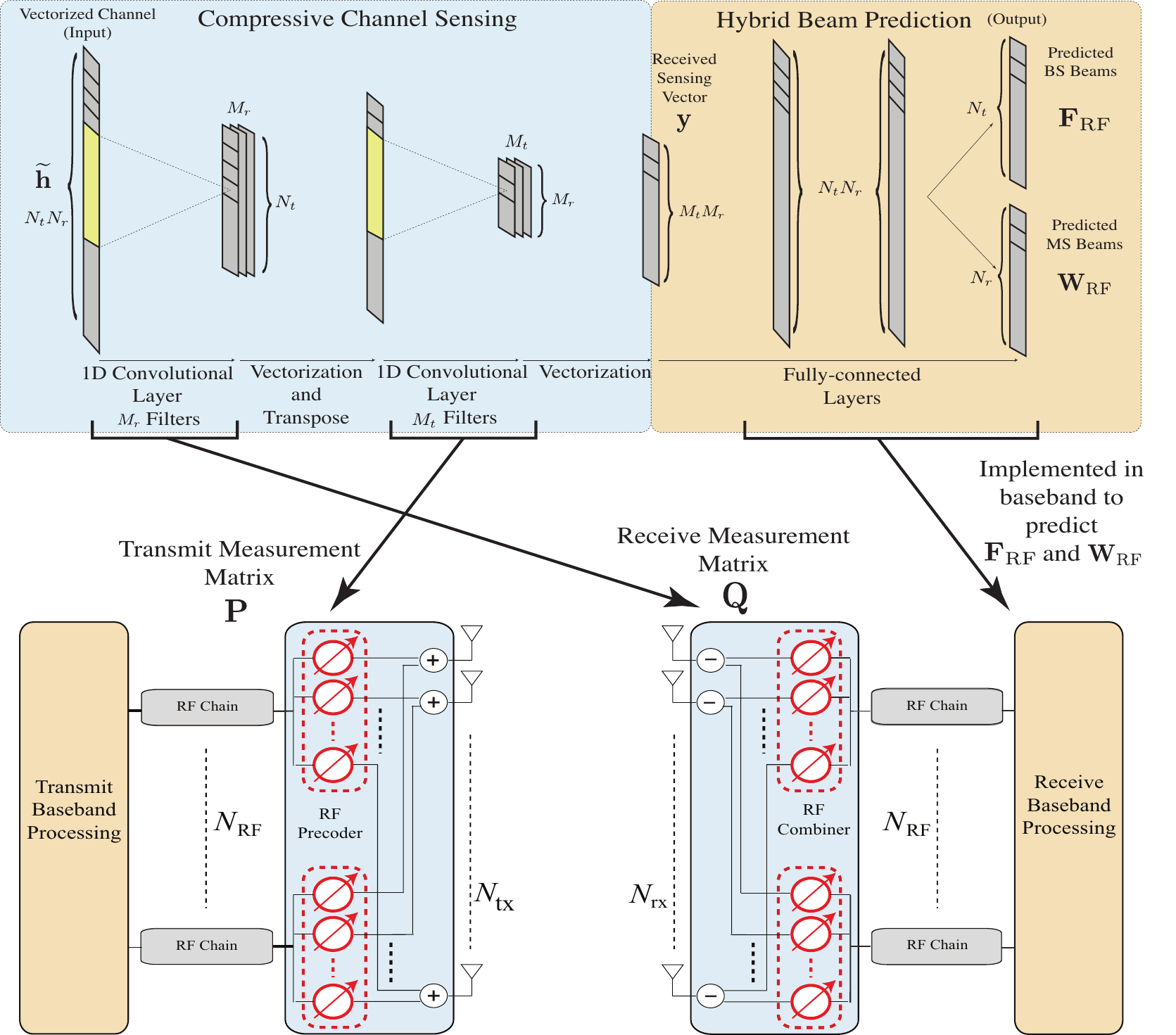}
	\caption{This figure illustrates the key idea of the proposed joint channel sensing and hybrid beam prediction approach. The complex weights of the trained channel encoding network are used as the weights of the transmit/receive measurement vectors that are implemented in the RF (or using hybrid architecture). The precoder network is implemented in the baseband. It takes the output of the channel measurement vectors (the receive vector) and use it to predict the best transmit/receive beamforming vectors.}
	\label{fig:Key_Idea}
\end{figure}

\subsection{mmWave Compressive Channel Sensing} \label{sec:sensing}

Leveraging the mmWave channel sparsity, \cite{Alkhateeb2014} proposed to leverage compressive sensing tools to sense and reconstruct the mmWave channels with hybrid analog/digital transceivers. In this section, we develop a novel neural network architecture that implements the compressive sensing formulation in \cite{Alkhateeb2014}, and allows the neural network model to  optimize the measurement vectors based on the surrounding environment and candidate user locations. 
More specifically, consider the system and channel models in \sref{sec:Sys_Model}. Let $\bP$ and $\bQ$ denote the $N_\mathrm{t} \times M_\mathrm{t}$ and $N_\mathrm{r} \times M_\mathrm{r}$ channel measurement matrices adopted by both the transmitter and receiver to sense the channel $\bH$, with $M_\mathrm{t}$ and $M_\mathrm{r}$ representing the number of transmit/receiver measurements. If the pilot symbols are equal to 1, then the received measurement matrix, $\bY$, can be written as  \cite{Alkhateeb2014}
\begin{equation}
\bY = \sqrt{P_\mathrm{T}} \bQ^\mathsf{H} \bH \bP + \bQ^\mathsf{H} \bV,
\end{equation}
 where $\left[\bV\right]_{m,n} \sim \cN_\mathbb{C}(0,\sigma_n^2)$ is the receive measurement noise. Now, when vectorizing this measurement matrix, $\bY$, we get 
 \begin{equation} \label{eq:sensing}
 \by=\sqrt{P_\mathrm{T}} \left(\bP^T \kron \bQ^\mathsf{H}\right) \bh+ \bv_q,
 \end{equation}
where $\by=\boldsymbol{\vec}\left(\bY\right), \bv=\boldsymbol{\vec}\left(\bQ^\mathsf{H} \bV\right)$ and $\bh=\boldsymbol{\vec}\left(\bH\right)$. 
In classical signal processing approaches (the do not leverage machine learning), the transmitters and receivers do not normally make use of previous observations and ,therefore, they do not have knowledge about the most promising spatial directions for the channel measurements. As a result, classical compressive channel sensing approaches typically adopt random measurement vectors \cite{Schniter2014,Alkhateeb2015}. Intuitively, however, given a certain environment (geometry, materials, user distribution, etc.), one would expect that these measurement vectors can be optimized based on the environment to improve the channel measurement performance. Next, we present a novel neural network architecture that mimics the joint transmitter/receiver channel sensing formulation in \eqref{eq:sensing} and allows for an environment-based optimization of the transmitter and receiver measurement vectors. 

\textbf{Network Architecture of `Channel Encoder':}
Looking at the channel sensing formulation in \eqref{eq:sensing}, we note that the product of the vectorized channel vector $\bh$ and the two measurement matrices can be emulated by inputing this channel vector into a neural network consisting of two consecutive convolutional layers, as shown in \figref{fig:multi_label}. In this architecture, the first convolutional layer employs  $M_\mathrm{r}$ kernels (filters). Each kernel has a size of $N_\mathrm{r}$ and a stride of $N_\mathrm{r}$, and represent one receive measurement vector. More specifically,  the weights of every kernel directly represent the entries of a receive measurement vector in $\bQ$. The output of the first convolutional layer has  $M_{r}$ feature maps. The matrix of these feature maps is then vectorized and fed in to the second convolutional layer. Similar to the first layer, the second convolutional layer consists of $M_\mathrm{t}$ kernels implementing the transmit measurement matrix $\bP$.  It is important to note here that since the transmitter/receiver measurement weights are generally complex, we adopt the complex-valued neural network implementation of the convolutional layers in \cite{Trabelsi2017}.  

It is important to note here that the end-to-end training of the auto-precoder (explained in \sref{subsec:Auto}) teaches this channel encoder in an unsupervised way to optimize its transmit/receive compressive channel measurement matrices (or kernel weights). Intuitively, this optimization adapts the measurement matrices to the surrounding environment and user distribution and focuses the sensing power on the promising spatial directions. After the model is trained, the kernels of the channel encoder will be directly employed as the measurement matrices by the transmitter and receiver to sense the unknown channels. The output of this channel measurement (the receive sensing vector $\by$) will be inputted to the precoder network in \figref{fig:multi_label}, which is the second section of the auto-precoder, to predict the RF beamforming/combining vectors.

\subsection{Hybrid beam predictions} \label{subsec:Precode}
Given the received channel measurement vector $\by$, we leverage deep neural networks to learn the direct mapping function from the received vector $\by$ and the  beamforming/combining vectors. For simplicity, we focus on predicting the RF beamforming/combining vectors $\bF_\mathrm{RF}$ and $\bW_\mathrm{RF}$. Note that finding these RF beamforming/combining vectors is what requires large training overhead in mmWave systems. Once the RF beamforming/combining vectors are designed, the low-dimensional effective channel $\bW_\mathrm{RF}^\mathsf{H} \bH \bF_\mathrm{RF}$ can be easily estimated and used to construct the beseband precoders and combiners. Further, since the RF beamforming/combining vectors are selected from the quantized codebooks $\boldsymbol{\mathcal{F}}, \boldsymbol{\mathcal{W}}$, we formulate the problem of predicting the indices of the RF beamforming/combining vectors as a multi-label classification problem. Next, we briefly explain the adopted network architecture for the hybrid beam prediction.

\textbf{Network Architecture of `Precoder':}
To predict the indices of the RF beamforming/combining vectors from the receive measurement vector $\by$, we propose the `precoder' neural network architecture in \figref{fig:multi_label}. This network consists of two fully connected layers and two output layers, and it is fed by the output of the `channel encoder' network described in  \sref{sec:sensing}. Each fully-connected layer is followed by Relu activation and batch normalization. The network has two output layers; one predicts the indices of the transmit beamforming vectors and the other one predicts the indices of the receive combining vectors. The dimensions of the two layers are equal to the cardinalities of the transmit and receive codebooks, $\boldsymbol{\mathcal{F}}, \boldsymbol{\mathcal{W}}$, i.e., the number of candidate beamforming/combining vectors. Note that the number of candidate beams is also the number of classes for the multi-label classification problem. 

The auto-precoder network is trained end-to-end in a Multi-Task Learning (MTL) manner \cite{ruder2017overview}, by considering the design of RF precoding and combining matrices of the hybrid architecture as two related tasks. This way, the network is trained to simultaneously optimize the two tasks which accounts for the dependence between the precoding and combining matrices. The MTL strategy enables us to solve two related problems using one neural network.

\subsection{Deep Learning Model Training and Prediction} \label{subsec:training}
As briefly highlighted in Sections \ref{subsec:Auto}-\ref{subsec:Precode}, the proposed auto-precoder network in \figref{fig:multi_label} is trained end-to-end as a multi-task learning  problem \cite{ruder2017overview}, which belongs to the supervised learning class. Essentially, we train the neural network based on a dataset of channel vectors and corresponding  RF beamforming/combining vectors of the hybrid architecture. The target RF precoding/combining matrices are calculated based on the near-optimal Gram-Schmidt hybrid precoding algorithm in \cite{Alkhateeb2016d}. Following \cite{Alkhateeb2018,Li2018b,Taha2019,Alrabeiah2019}, the channel vectors are globally normalized such as the largest absolute value of the channel elements equals one.  The labels for the transmit beamforming vectors (and similarly for the receive combining vectors) are modeled as $N_\mathrm{t}^\mathrm{RF}$-hot vectors, with ones at the locations that correspond to the indices of the target RF beamforming codewords (from the codebook $\boldsymbol{\mathcal{\bF}}$). 

\textbf{For the loss function}, we use the binary cross entropy for the multi-label classification that corresponds to each task (i.e., to the precoding and combining tasks). The total loss function is the arithmetic mean of the binary cross entropies of the two tasks since they are equally important for the entire hybrid precoding/combing design objective. To evaluation the prediction performance, we calculate the sample-wise accuracy of the predicted indices. Let $y_{\rm true}^{(i)}$ denote the set of label indices and $y_{\rm pred}^{(i)}$ denote the set of predicted indices. Then, the accuracy of sample $i$ is defined as
\begin{align}\label{eqn:precision}
a_{i}=\frac{\big|y_{\rm true}^{(i)} \cap y_{\rm pred}^{(i)}\big|}{\big|y_{\rm true}^{(i)}\big|},
\end{align}
where $\cap$ represents the intersection of two sets and $\big|\cdot\big|$ is the cardinality of a set. Finally, the sample-wise accuracy is defined as
\begin{align}\label{eqn:avgprec}
a=\frac{\sum_{i=1}^{N}a_{i}}{N_\mathrm{samples}},
\end{align}
where $N_\mathrm{samples}$ is number of samples. Note that since the number of labels for each channel is a fixed number for both true labels and predicted labels, the two widely-adopted performance metrics in the multi-label classification problems, namely precision and recall, reduce to (\ref{eqn:precision}).

During this training process, the neural network architecture in \figref{fig:multi_label}  achieves two joint objectives: (i) It optimizes the transmitter/receiver measurement vectors (which are the weights of the two convolutional layers in the channel encoder network) in an unsupervised way to focus the sensing power on the most promising directions and (ii) it learns how to predict the RF beamforming/combining vectors of the hybrid architecture directly from the channel measurement vectors through the precoder network. Thanks to this design that leverages the prior observations in designing the channel measurement/sensing vectors and hybrid precoding/combining matrices, the proposed approach significantly reduces the required training overhead to design the hybrid precoding/combining matrices, as will be shown in the following section. 

\section{Simulation Results} \label{sec:Results}
In this section, we evaluate the performance of the proposed deep-learning based direct hybrid precoding design approach using realistic 3D ray-tracing simulations.

\subsection{Dataset and Training}\label{sec:dataset}
Our dataset is generated using the publicly-available generic DeepMIMO \cite{DeepMIMO} dataset with the parameters summarized in Table \ref{tab:param_set}. More specifically, we consider BS 4 in the street-level outdoor scenario `O1' communicating with the mobile users from row R1200 to R1500 with an uplink setup. For simplicity, both the transmitter (mobile users) and receiver (base station) are assumed to employ $64$  antennas with $3$ RF chains each.  For every user, we first construct the channel matrix using the DeepMIMO dataset generator. Then, random noise is added to the channel matrix. The noise power is calculated based on a bandwidth of $0.5$GHz and receive noise figure of $5$dB. The channel is also adopted to construct the target RF precoding/combining matrices of the hybrid architecture using the near-optimal Gram-Schmidt based hybrid precoding algorithm in \cite{Alkhateeb2016d}. the pair of the noisy channel and the corresponding codebook indices  of the RF precoders/combiners is then considered as one data point in the dataset. 

\begin{table}[!t]
	\caption{The adopted DeepMIMO dataset parameters}
	\begin{center}
		\begin{tabular}{|c|c|}
			\hline
			DeepMIMO Dataset Parameters	&Value	\\
			\hline
			Activate BS	&4  \\
			\hline
			Activate users &From row R1200 to R1500  \\        
			\hline
			Number of BS Antennas &$M_x=1$, $M_y=64$, $M_z=1$   \\        
			\hline
			Number of User Antennas &$M_x=1$, $M_y=64$, $M_z=1$   \\        
			\hline
			Antenna spacing (in wavelength) &$0.5$ \\        
			\hline
			System bandwidth (in GHz) &$0.5$\\        
			\hline
			Number of OFDM subcarriers &1024 \\        
			\hline
			OFDM sampling factor &1 \\        
			\hline
			OFDM limit &1 \\        
			\hline
			Number of paths &3 \\        
			\hline
		\end{tabular}
	\end{center}
	
	\label{tab:param_set}
\end{table}

The constructed dataset is used to train the proposed auto-precoder neural network adopting the cross-entropy based loss function defined in \sref{subsec:training}. Our model is implemented using the Keras libraries \cite{Branchaud-Charron} with a Theano \cite{Bergstra2011} backend. We use the Adam optimizer with momentum $0.5$, a batch size of $512$, and a $0.005$ learning rate.

\subsection{Achievable Rates}
To evaluate our deep-learning based direct hybrid precoding solution, we calculate the achievable rate using the predicted precoding/combining indices and compare it with the optimal rate. More specifically, for a given noisy channel measurement, we use the proposed auto-precoder neural network model in \sref{sec:hybrid} to predict the indices of the RF precoding/combining matrices of the hybrid architecture. Then, adopting the baseband precoding/combining design in \eqref{eq:baseband1}-\eqref{eq:baseband2} we calculate the achievable rate as defined in \eqref{eq:rate}. This rate is then compared to the optimal rate that is achieved when the precoding/combining matrices are calculated as described in \sref{sec:dataset} with perfect channel knowledge.

\begin{figure}
	\centering
	\includegraphics[width=1\columnwidth]{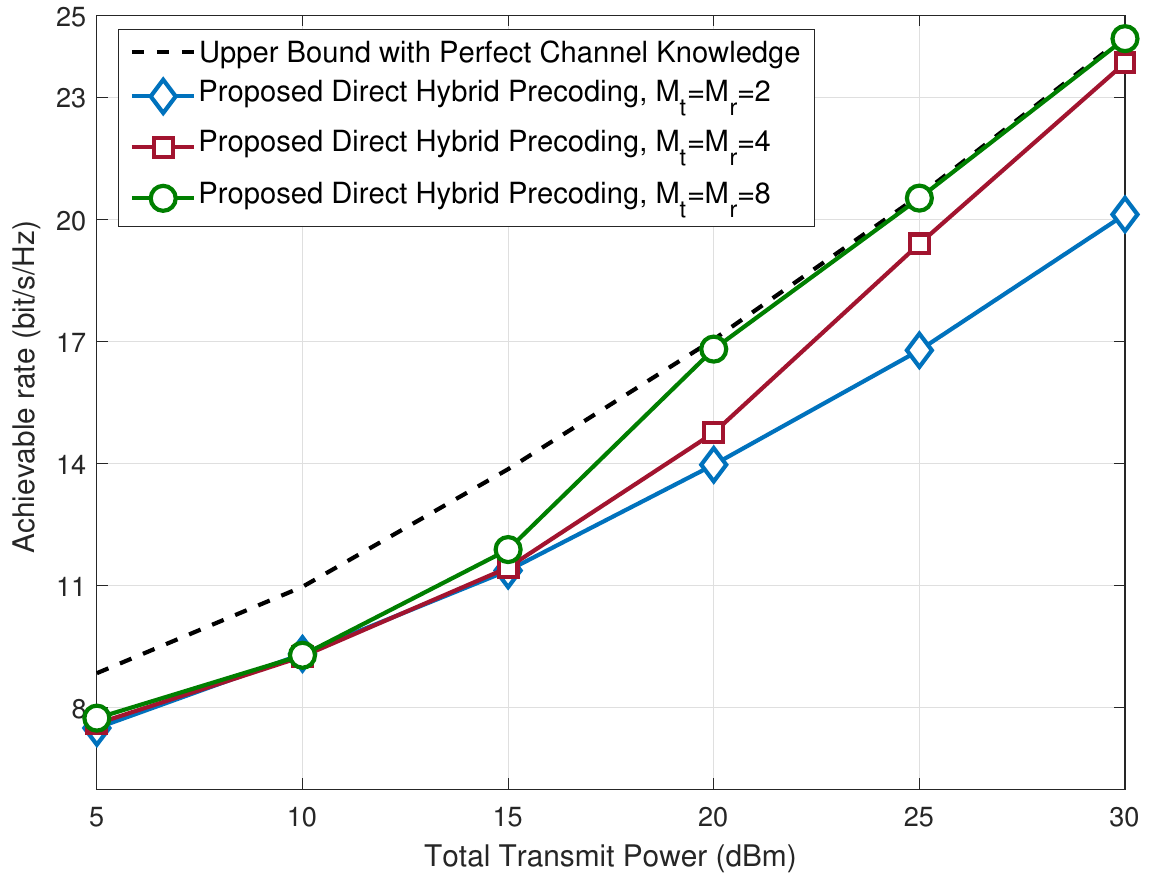}
	\caption{The achievable rates of the proposed deep-leaning based joint channel sensing and hybrid precoding approach for different values of total transmit power. $P_\mathrm{T}$. These rates are also compared to the upper bound that designs the hybrid precoding/combining matrices assuming perfect channel knowledge. This figure illustrates that the proposed solution can achieve near-optimal data rates while significantly reducing the training overhead, i.e. requiring a small number of channel measurements, $M_\mathrm{t}, M_\mathrm{r}$). }
	\label{fig:rate_SNR}
\end{figure}

Fig. \ref{fig:rate_SNR} illustrates the achievable rate of the proposed deep-learning based blind hybrid precoding approach and its upper bound versus different values of total transmit power. Further, to draw some insights into the required training overhead, we plot the achievable rate of the proposed direct hybrid precoding approach for three different values of the channel measurements,  $M_\mathrm{t}=M_\mathrm{r}=2, 4, 8$. This figure adopts the system and channel models described in \sref{sec:dataset} where both the transmitter (mobile user) and receiver (base station) employ a hybrid transceiver architecture with $64$-element ULA and $3$ RF chains. First, Fig. \ref{fig:rate_SNR} shows that the performance of the proposed deep-learning based direct/blind hybrid precoding solution approaches the upper bound (which assumes perfect channel knowledge) at reasonable values of the transmit power.  Further, and more interestingly, this figure illustrates the significant reduction in the required training overhead compared to classical (non-machine learning) solutions. For example, with only $4$ channel measurements at both the transmitter and receiver, i.e., a total of $4 \times 4 = 16$ pilots, the proposed deep-learning based channel sensing/precoding solution achieves nearly the same spectral efficiency of the upper bound. This is instead of the $64 \times 64 = 4096$ pilots required for exhaustive search and almost one tenth of that, $\sim 300-400$, pilots needed by compressive sensing solutions \cite{Alkhateeb2015}. \textbf{This significant reduction in the training overhead is thanks to the proposed deep learning model which leverages the prior observations to optimize the channel sensing based on the environment/user locations and  to learn the mapping from the receive channel sensing vector and the optimal precoders/combiners. }

\subsection{Sample-Wise Accuracy}
In  addition to the achievable rate (which is the main objective of this work), we also evaluate the performance of the proposed deep learning model using the sample-wise accuracy defined in \sref{subsec:training}. The sample-wise accuracy evaluates the ability of the deep learning model in predicting the correct set of the RF beamforming/combining vectors for the hybrid architecture. In Table \ref{tab:acc_vs_mea}, we adopt the same system and channel models considered in \figref{fig:rate_SNR} and summarize the sample-wise accuracy of the transmit beamforming and receive combining beams for different values of total transmit power $P_\mathrm{T}$. Similar to \figref{fig:rate_SNR}, Table \ref{tab:acc_vs_mea} shows with only 4 or 8 channel measurements, the  sample-wise accuracy of predicting the exact 3 transmit beams and $3$ receive beams reaches nearly $90\%$. It is worth noting here that the accuracy for the transmit and receive predicted beams are almost the same since we treat them as equally important tasks when training the auto-precoder neural network model.

\begin{table}[!t]
	\caption{Beam Indices prediction accuracy versus SNR}
	\begin{center}
		\begin{tabular}{c|c|c|c|c|c|c}
			\hline
			$P_T$ (dBm)	    & 5	    &10	    &15	      &20	  &25    &30\\
			\hline
			Tx acc. ($M_\mathrm{t}$=$M_\mathrm{r}$=2)	&0.56    &0.63    &0.71    &0.72    &0.73    &0.74\\
			\hline
			Rx acc. ($M_\mathrm{t}$=$M_\mathrm{r}$=2)  &0.55    &0.63   &0.70    &0.71    &0.72  &0.72\\        
			\hline
			Tx acc. ($M_\mathrm{t}$=$M_\mathrm{r}$=4)	&0.67   &0.69    &0.72    &0.77    &0.85    &0.88\\
			\hline
			Rx acc. ($M_\mathrm{t}$=$M_\mathrm{r}$=4)  &0.66    &0.69   &0.71   &0.78    &0.85    &0.88\\        
			\hline
			Tx acc. ($M_\mathrm{t}$=$M_\mathrm{r}$=8)	&0.68    &0.70   &0.73    &0.89    &0.91    &0.91\\
			\hline
			Rx acc. ($M_\mathrm{t}$=$M_\mathrm{r}$=8)   &0.61    &0.69   &0.73    &0.89    &0.91    &0.92\\        
			\hline
		\end{tabular}
	\end{center}
	
	\label{tab:acc_vs_mea}
\end{table}

\section{Conclusion}
In this paper, we developed a neural network architecture and a deep-learning approach for joint channel sensing and hybrid beamforming design in mmWave massive MIMO systems. The proposed neural network, that we called an auto-precoder, has two components: (i) The channel encoder which learns how to optimize the channel sensing vectors to focus the sensing power on the promising directions, and (ii) the precoder network which learns how to predict the RF beamforming/combining vectors of the hybrid architecture directly from the  received sensing vector.  Thanks to the specific design of the channel encoder that uses complex-valued neural networks and accounts for the constraint on the RF chains, the trained weights of the channel encoder are directly used as channel measurement vectors in the prediction phase. Simulation results showed that the proposed deep-learning based solution can successfully predict the hybrid beamforming vectors that achieve near-optimal data rates while requiring negligible training overhead compared to exhaustive search and classical compressive sensing solutions. 

\balance

\end{document}